# A polynomial time (heuristic) SAT algorithm

Charles Sauerbier

ABSTRACT — [Note: See addendum prefacing this paper, as a hole has been determined to exist in the algorithm as presented in this paper.] An algorithm of two parts is presented that determines existence of and instance of an assignment satisfying of instances of SAT. The algorithm employs an unconventional approach premised on set theory, which does not use search or resolution, to partition the set of all assignments into non-satisfying and satisfying assignments. The algorithm, correctness, and time and space complexity proofs are given.

KEY TERMS — **Algorithms, complexity, computation theory, satisfiability, set theory.**

ADDENDUM

A hole has been found in the algorithm as presented, where an eleventh hour change admits a path inconsistency resulting in false affirmation of existence of a solution at the end of Part A of the algorithm. This results from cyclic closure of paths against a root other than that supporting the path as its origin. Existence of a path can still be established using the algorithm where Part B is used to search the reduced space represented in the results of Part A. However, to fully and completely effect this check is to nullify any benefit of having performed Part A in alternative to the motivating basis from which Part A was derived. A reversion back to the more primitive and computationally expensive in the worst case is being composed. For now the algorithm as presented can be said to be no better than a polynomial time heuristic SAT algorithm. The algorithm has not been found to give false negatives. The underlying mathematical premises motivating the algorithm do not support the algorithm giving false negatives.

In making the change the impact to consistency was not given sufficient consideration. Reversion to the mechanism from which the version given here derives results in a increase in the exponent value of 3 for worst-case performance of the algorithm, though best-case performance is potentially improved by in reduction of the exponent value for some instances of SAT.

I. INTRODUCTION

Since Stephen Cook first proposed the "satisfiability" (SAT) problem [1], an open question has existed in computer science: *Does there exist a polynomial time algorithm for solution of this problem?* [2] gives discussion of the problem in some detail. The search for a resolution to this question found its base in the works of Davis, et al, [3][4]. These early works, predating [1], formed the basis for the commonly employed techniques of backtracking search [3][5] and resolution [4][6][7] used by current SAT solvers; see [5]. The ongoing interest in the satisfiability problem has also produced a varied array of heuristic techniques in effort to derive solutions.

A radically different approach is taken in the presented algorithm of this paper. The algorithm does not employ any form of search. The algorithm also does not employ any form of variable/clause resolution. The algorithm is premised on a set theoretic view of the SAT problem. The problem is reconstructed in terms of set theory as two sets containing as elements sets of variables in a given instance, wherein the actual variables are abstracted away and the underlying patterns in clauses are then used to determine the sets of inadmissible assignments. Admissibility of assignment patterns within the context of both sets being then determined, the effect of the algorithm is the partitioning of the set of all possible assignments into two disjoint sets: non-satisfying and satisfying. This partitioning is possible in polynomial time in consequence of the finality of inadmissibility of assignments to variables in the instance of determination. The finality of such comes in consequence of the fundamental definition of satisfying assignments and the necessary requirement that exactly one assignment to any variable can exist in any admissible assignment.

The algorithm presented determines whether there exists an assignment satisfying of an instance of SAT by determining if there exists a non-empty set of admissible assignments. The algorithm is premised on set partitioning. It is shown to have deterministic polynomial time and space complexity, with upper bound worst case time complexity of $O(n^8)$ and space complexity of $O(n^3)$.

The question of a solution based on other techniques, such as search and resolution, is left as an NP-Complete problem. Problems not amenable in polynomial time to the approach taken here also remain NP-Complete. It is conjectured in [2] that such problems do exist.

## II. APPROACH

The SAT problem asks: *Does there exist an assignment to a set of variables such that an instance of a SAT expression evaluates to true*? Traditionally the SAT problem has been approached from the perspective of searching for an instance of a satisfying assignment in the space of all possible assignments. For any instance of SAT such assignment space is $2^n$, in the number of variables n. This makes an exhaustive search of all possible assignments to an instance of the problem plausible for only the smallest values of n.

The approach taken here is based on the premise that for an assignment satisfying an instance of SAT to exist, the set of all satisfying assignments must not be an empty set. The question is therefore restated: *Does there exist a non-empty set of satisfying assignments to any given instance of SAT*? Determining the set of all satisfying assignments is equivalent to determining the set of all non-satisfying assignments, as the two sets form a disjoint partition of the set of all assignments for any instance of SAT. In set notation we have $A = N \cup S$ and $N \cap S = \emptyset$; where A is the set of all assignments, N the set of non-satisfying assignments, and S the set of satisfying assignments. It follows that if one can show N is determinable in polynomial time then S is also determinable in polynomial time in consequence of S being the complement of N in A (i.e.: $A \backslash N = S$).

The question then becomes one of means to effect the reduction of the set of all possible assignments to the set of all satisfying assignments. Since it is commonly known that all instances of SAT are polynomial time reducible to 3SAT [2], the means comes in consequence of three attributes of 3SAT:

1. The set of clauses of any instance of 3SAT may be partitioned into sets such that each set contains all the clauses in the instance that are composed of the same 3-tuple of variables, independent of negation (clausal partition), and;
2. Any assignment satisfying of SAT must contain for each element of the clausal partition exactly one

assignment to the 3-tuple of variables defining of the respective clausal partition element that is satisfying of all clauses in the element, and;

3. Any assignment satisfying of SAT must make exactly one assignment to the variables in common between any elements of the clausal partition, and thus must admit a mutually satisfying assignment in all elements of the clausal partition having the variables in common.

Note that the formal definition of clausal partition can be found in section III (Definitions) below.

Search for an assignment satisfying of an instance of SAT is one means to determine if an assignment to the 3-tuple of variables defining of an element in the clausal partition is contained in an assignment satisfying of the SAT instance. Solutions to SAT instances have conventionally been sought by searching for such an assignment by consideration of assignment to individual clauses or in some cases individual variables.

It is obvious, from the three conditions stated above, that where an assignment to the 3-tuple of variables defining of a clausal partition element is at any time determined to be inadmissible as satisfying such remains an immutable fact of all assignments satisfying of the instance of SAT. This is to say that no assignment satisfying of the instance may contain that assignment to the variables. This follows from (2) and (3) above. It is important to note here that a determination of any assignment to a specific set of variables being non-satisfying is valid only on the set of variables in whole, and not individually.

As all instances of SAT are transformable into an instance of 3SAT, the satisfaction problem is considered in this paper in terms of 3SAT. For discussion an instance of 3SAT over the following set of clauses will be presumed in example to aid in explanation of meaning of terms used in this paper:

$$C = \{\ c_0 = \alpha_1 \vee \neg\alpha_2 \vee \neg\alpha_3,\ \ c_1 = \neg\alpha_1 \vee \alpha_2 \vee \neg\alpha_3,\ \ c_2 = \neg\alpha_1 \vee \neg\alpha_2 \vee \alpha_3,$$
$$c_3 = \alpha_2 \vee \neg\alpha_3 \vee \neg\alpha_4,\ \ c_4 = \neg\alpha_2 \vee \neg\alpha_3 \vee \alpha_4,\ \ c_5 = \neg\alpha_2 \vee \neg\alpha_3 \vee \neg\alpha_4,$$
$$c_6 = \neg\alpha_3 \vee \alpha_4 \vee \alpha_5,\ \ c_6 = \neg\alpha_3 \vee \neg\alpha_4 \vee \alpha_5,\ \ c_7 = \neg\alpha_3 \vee \neg\alpha_4 \vee \neg\alpha_5\ \}.$$

These clauses can then be partitioned into what is termed in this paper as the clausal partition as follows:

$$D = \{\ d_1 = \{c_0 = \alpha_1 \vee \neg\alpha_2 \vee \neg\alpha_3,\ c_1 = \neg\alpha_1 \vee \alpha_2 \vee \neg\alpha_3,\ c_2 = \neg\alpha_1 \vee \neg\alpha_2 \vee \alpha_3\ \},$$
$$d_2 = \{c_3 = \alpha_2 \vee \neg\alpha_3 \vee \neg\alpha_4,\ c_4 = \neg\alpha_2 \vee \neg\alpha_3 \vee \alpha_4,\ c_5 = \neg\alpha_2 \vee \neg\alpha_3 \vee \neg\alpha_4\ \},$$
$$d_3 = \{c_6 = \neg\alpha_3 \vee \alpha_4 \vee \alpha_5,\ \ \ \ c_6 = \neg\alpha_3 \vee \neg\alpha_4 \vee \alpha_5,\ c_7 = \neg\alpha_3 \vee \neg\alpha_4 \vee \neg\alpha_5\ \}\ \}.$$

The problem is approached in this paper from a perspective of set theory. Any assignment satisfying of SAT must contain exactly one assignment to any variable and thus for any 3-tuple of variables. It follows then that any assignment not satisfying of all clauses in the set of such defined by a clausal partition element of the SAT instance is inadmissible in any assignment satisfying of the instance of SAT by definition. The compliment of the set of inadmissible assignments for a clausal partition element is the set of admissible assignments for the set of clauses. Where one considers only the set of clauses in the clausal element the compliment of the set of inadmissible assignments is the set of admissible assignments for that set of clauses and only that set of clauses. The later is the essence of (2) above and its expression in the context of the algorithm.

The clausal partition elements, however, define a second set of sets: The set containing as elements the set of variables in the intersection of the 3-tuples of variables defining of clausal partition elements. For the example clausal partition this set is the set $\{\{\alpha_2, \alpha_3\}, \{\alpha_3, \alpha_4\}\}$. The assignment to the variables in each of these sets must be mutually admissible by all clausal partition elements for any such assignment to be contained in an assignment satisfying of SAT, for the same reason such is required of assignments within each clausal partition element: Only a single assignment to the variables may exist in any assignment to the SAT instance. The algorithm determines the mutually admissible assignments to the elements of this set by means of successive application of a defined implication operator.

The implication operator propagates constraints on assignment to variables by means of imposition of constraint from one clausal partition element into another. Constraints are defined to be exclusory of assignments. For the discussion example the assignment constraint set for $d_1$ is the set $\{(F, T, T), (T, F, T), (T, T, F)\}$, for $d_2$ is the set $\{(F, T, T), (T, T, F), (T, T, T)\}$, and for $d_3$ is the set $\{(T, F, F), (T, T, F), (T, T, T)\}$. The set of constraints to be propagated by imposition is determined by evaluating whether all possible assignments to a variable subset of the 3-tuple defining the clausal partition element are in the excluded set for the element. The discussion example clausal partition element $d_2$ contains exclusions of both assignments for $\alpha_3 = T$ and $\alpha_4 = T$ (i.e.: (T, T, T) and (F, T, T)). This implies that no assignment having (T, T) for variables $\{\alpha_3, \alpha_4\}$ can ever be an assignment satisfying of the associated instance of SAT. Consequently, the implication operator being applied to clausal partition elements $d_2$ and $d_3$, imposing $d_2$ on to $d_3$, would impose on $d_3$ the constraints $\{(T, T, T), (F, T, T)\}$ as inadmissible. Since the intersection of 3-tuples defining of a clausal partition element are always considered as a set, an intersection on any one variable of $d_2$ would not impose the constraint on assignment of $\alpha_3 = T$ or $\alpha_4 = T$ individually or in their individual combination with another variable (e.g.: $\{\alpha_2, \alpha_3\}$) based on the constraint thus determined for $\{\alpha_3, \alpha_4\}$.

As a point of clarification, the clausal partition elements are always considered in terms of a 3-tuple of variables. Obviously this implies the possible existence of clausal partition elements containing fewer than 3 distinct variables. The algorithm depending on intersection of variable sets, no loss of generality is had to proofs where sets are considered to always have three distinct variables.

There being at most three distinct variables in each clause the cardinality of each set of clauses is assured to be less than or equal to eight; with each set of clauses having exactly eight possible assignments. For each element of the clausal partition at most one of the eight possible assignments can exist in any one instance of an assignment satisfying of an instance of SAT. A possible assignment in any one element of the clausal partition may be contained in any or none of the assignments satisfying of an instance of SAT. Whether an admissible assignment in a clausal partition element is contained in any assignment satisfying of an instance of SAT is a consequence of the relation between elements of the clausal partition that results from having variables in common. It is the relation between elements of the clausal partition through variables in common that allows the means used to partition the set of all possible assignments into the set of non-satisfying and the set of satisfying assignments in the algorithm presented.

The SAT problem in whole is transformed into a graph representation (instance graph) on the clausal partition, where vertices represent clausal partition elements and edges represent the existence of a non-empty intersection of the defining 3-tuble of variables of associated clausal partition elements. The algorithm propagates assignment constraints using the instance graph and implication operator.

The instance graph in combination with the finite upper bound of 8 on the number of possible clause in and assignments to each clausal partition element allows deterministic expression of bounds on the time and space complexity of the algorithm that are polynomial in the number of variables of the problem instance. The algorithm propagates assignment constraints using the instance graph and implication operator until a steady state is achieved. That is to say: Until no further imposition of constraints happens in the primary loop of the algorithm. Since the implication operator is simply eliminating assignments as inadmissible from the set of all possible assignments for each clausal partition element, the system must reach a steady state at or prior to where all clausal partition elements have all possible assignments eliminated.

### III. Definitions

The following definitions are made and used throughout the subsequent text of the paper. Where some may be trivially obvious or commonly known such are yet given here for their potential utility in brevity later.

**Definition:** *Disjunctive Clause* – Given any set of variables U taking values from the set {True, False} (or by equivalence {0, 1}), any clause over such variables wherein the variables hold relation by no operator other than the disjunctive operator is defined to be a disjunctive clause. (i.e.: $c = \omega_1 \vee \omega_2 \vee \ldots \vee \omega_n$, where $\omega_i = u_i$ or $\omega_i = \neg u_i$ and $u_i \in U$).

**Definition:** *varset(c) operator* (clauses) – Given any set of variables U, let c be a clause over some set of variables $u_i \in U$, *varset(c)* is then the set of all variables in c.

**Definition:** *satset(C) operator* (clause set) – Given any set of variables U, let C be any set of clauses, $c_i$, on variables of U; *satset(C)* is then the set of admissible assignments of C such that all $c_i \in C$ are satisfied.

**Definition:** *varcom($C_i$, $C_j$) operator* (clause set) – Given any set of variables U, let $C_i$ and $C_j$ be any two set of clauses on variables of U; varcom($C_i$, $C_j$) = {varset($C_i$) $\cap$ varset($C_j$)} is then the set of variables in common between $C_i$ and $C_j$.

**Definition:** *asgspc(U) operator* (assignment space) – Given some set of variables, asgspc(U) is the set of all possible assignments of values to variables in U.

**Definition:** *Reductive Property* – A relation (operator) on sets has the property of being "reductive" if and only if the relation reduces the set(s) on which it operates by zero or more elements, and cannot increase in number the elements of set(s) on which it operates.

**Definition:** *Strictly Reductive Property* – A relation (operator) on sets has the property of being "strictly reductive" if and only if the relation reduces the set(s) on which it operates by one or more elements, and cannot increase in number the elements of set(s) on which it operates.

**Definition:** *Non-Reductive Property* – A relation (operator) on sets has the property of being "non-reductive" if and only if the relation increases in number the elements of set(s) on which it operates by zero or more elements, and cannot reduce in number the elements of set(s) on which it operates.

**Definition:** *Implication Operator* $\nabla$ (on sets of clauses) – Given a set of clauses A such that $|varset(a_i)| = |varset(a_j)| = n \ \forall \ a_i, a_j \in A$, and, $|\{ \cap varset(a_i) \ \forall \ a_i \in A \}| = n$, and; a set of clauses B similarly defined: define $\nabla: A \to B$ such that where $varcom(A, B) \neq \emptyset$, the variables in common, A imposes constraint on B to the exclusion of assignments, as inadmissible, any having assignments of values for the common variables that are excluded as inadmissible in the context of A.

**Definition:** *Implication Operator* $_2\nabla$ (bi-directional on sets of clauses) – Given a set of clauses A such that $|varset(a_i)| = |varset(a_j)| = n \ \forall \ a_i, a_j \in A$, and, $|\{ \cap varset(a_i) \ \forall \ a_i \in A \}| = n$, and; a set of clauses B similarly defined: define $\nabla: A \leftrightarrow B$ to be the successive alternating application of the implication operator in unidirectional manner $A \to B$ and $B \to A$ until a steady state is achieved.

**Definition:** *3SAT* – Let U be a finite set of variables taking values in the set {True, False}. Let C be a collection of disjunctive clauses on the variables of U such that $|c| = 3 \ \forall \ c \in C$. Let E be a conjunction of all clauses in C (i.e.: $E = c_1 \wedge c_2 \wedge \ldots \wedge c_n$). Then $E = (U, C)$ is an instance of 3SAT.

**Definition:** *Satisfiable* (clauses) – A clause formed as the disjunction of logical variables taking values from {True, False} is satisfiable for any assignment of values such that at least one variable in the clause evaluates to true.

**Definition:** *Satisfiable* (SAT) – An instance of SAT is satisfiable if and only if there exists at least one assignment of values to each variable in U so that all clauses in C are satisfiable.

**Definition:** *Satisfying Assignment* (Clause Set) – An admissible assignment of values to the variables of a set of disjunctive clauses for which all clauses of the set evaluate to TRUE.

**Definition:** *Satisfying Assignment* (Expression) – An admissible assignment of values to the variables of a conjunctive expression on some set of disjunctive clauses for which the expression evaluates to TRUE.

**Definition:** *$\emptyset$ Satisfying Assignment* – Let C be a set of disjunctive clauses over n variables such that $|c_i| = |c_j| = n \ \forall \ c_i, c_j \in C$. Let $\alpha_1, \alpha_2, \ldots \alpha_n$ be the variables of the clauses of C. A variable $\alpha_i$ in C has a $\emptyset$ satisfying assignment if and only if no assignment of values to the variable can satisfy C, and no admissible assignment of values to another variable of C may satisfy C.

**Definition:** *Clausal Partition, D* – Let C be a set of clauses in 3SAT form over a set of variables U. Define the clausal partition $D = \{d_1, d_2, \ldots d_n\}$, to be a partition of C such that each $d_i \in D$ is associated with a unique 3-tuple of variables from U, and; $d_i$ contains all $c_k \in C$ where $c_k$ is composed of only the 3-tuple of variables defining $d_i$, without consideration of negation, and; for all $c_k \in C$ there exists one and only one $d_i$ containing $c_k$.

**Definition:** *SAT Instance Graph, G* – Given some instance, E = (U, C), of SAT in 3SAT form and the clausal partition, D, define G = (V, β) to be the graph where V = {$v_i$ | ∀ $d_i$∈ D ∃ $v_i$ uniquely associated with $d_i$}, thus |V| = |D|, and; β = {($v_i$, $v_j$) | 0 < |varset($v_i$) ∩ varset($v_j$)| < 3; $v_i$ ≠ $v_j$ }.

## IV. VARIABLE VALUES

SAT is by definition about nothing more than assignment to variables a value from the set {T, F} under constraint of context within a logical expression. The approach taken here is based on sets, and as such it follows that the set of admissible assignment values is of some concern.

**Proposition 1:** *Admissible assignment values set for variables of clauses in an instance of SAT*

Let E = (U, C) be a conjunctive expression on a set of disjunctive clauses C in a set of variables U taking values from the set {F, T}. By definition each variable has an associated set of possible assignments. By consequence of common variable relations in the set of clauses C, each variable in U has a constrained subset of all possible assignments that are admissible as participant in the set of assignments satisfying of E. The set of all possible assignment values being taken from the power set over the set from which individual variable values are taken unioned with the empty set (i.e.: {{True, False}, {True}, {False}, ∅}). The proof follows from the definitions of SAT and satisfiable.

**Proposition 2:** *Empty satisfying set as necessary consequence to empty admissible assignment.*

Let E = (U, C) be a conjunctive expression on a set of disjunctive clauses C in a set of variables U taking values from the set {F, T}. If any variable in U has the set ∅ as an admissible set of assignments then the set of assignments satisfying E is the set ∅.

**Proof:** Let $\alpha_1, \alpha_2, \ldots, \alpha_n$ be the variables of E. If the set of variables {$a_j$ | $a_j$ ∈ ({$\alpha_1, \alpha_2, \ldots, \alpha_n$} − {$\alpha_i$})} contained a admissible assignment satisfying of E, then the set of assignments satisfying E is independent of $\alpha_i$; therefore $\alpha_I$ = {T, F} ≠ ∅. It follows then that if the admissible assignment of any variable $\alpha_i$ = ∅ the set of variables {$a_j$ | $a_j$ ∈ ({$\alpha_1, \alpha_2, \ldots, \alpha_n$} − {$\alpha_i$})} has no admissible assignment such that E can be satisfied. It therefore follows directly from the definition that the set of assignments satisfying of E is the set ∅.

## V. CLAUSAL PARTITION

The SAT problem can be partitioning into two set problems: one on independent variables within clauses, and, one on dependent (common) variables between clauses. The later is the crux of the problem in solving SAT, as well as the key to solving it. The determination of admissible assignments to the set of common variable sets is dependent on the simultaneous solution of admissible assignments of each set of clauses containing the variables in common. The problem of simultaneous determination of the two sets is handled in the algorithm by a combination of the structure of the instance graph and application of the implication operator to elements of the clausal partition. The clausal partition elements define the only object to which the implication operator is applied in the algorithm.

A clausal partition is derived from the set of clauses, C, in the instance, E = (U, C), of SAT in 3SAT form by partitioning C into a set of subsets where each subset contains all clauses in C having the same set (i.e.: defined by the same 3-tuple) of variables, ignoring whether the variables are non-negated or negated in the clause. Except where clauses contain two instances or three instances of one variable in U, all clauses in any element in the clausal partition otherwise have three distinct variables in common.

Clausal partition elements are represented in the algorithm by objects that are one of its principal data structures. The attributes of the representational object of clausal partition elements are a 3-tuple of variables, the set of patterns on the variables for which clauses exist in C, and a set of constraints on the set of possible assignments to the variables defining the set of clauses. A clausal partition element's admissible assignment set is derived by taking the compliment of its set of constraints; the later defining the set of inadmissible assignments for the element.

The algorithm is not concerned with actual variables for most of its function. The algorithm is concerned with the set of patterns on 3-tuples for which clauses exist in the SAT instance, in determining admissible assignment. It is not the variables, but rather the existence of patterns defined by the clauses over the set of variables that determines the initial (explicit) constraints on assignments satisfying of the set of clauses, independent of their context within E. Subsequent implication of assignment constraints for variables in common is dependent on the patterns defined by constraints present in each respective element of the clausal partition. The actual variables have relevance to the algorithm only in determining the clausal partition, in determining which elements of that partition have variables in common, and in determining which variables are the common variables between any two clausal partition elements.

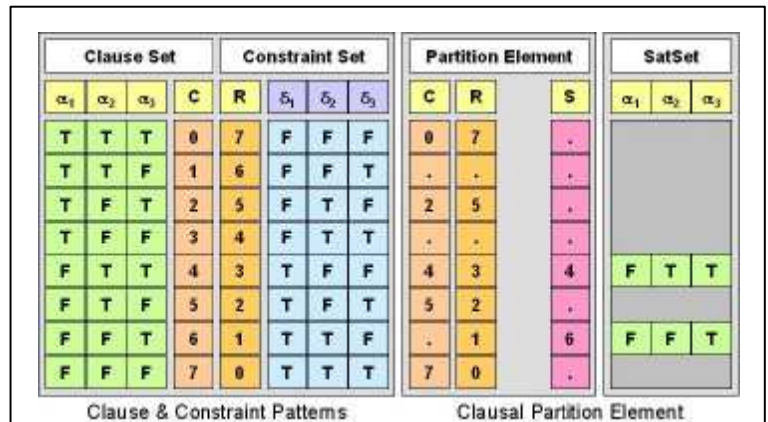

Fig. 1: Left panel gives set sets of patterns existing on 3-tuples of variables in 3SAT for clauses and constraints, and associated bit-vector representations of each used by algorithm. Right panel illustrates a bit-vector representation of clausal partition element containing clause, constrain, and satisfying-assignment sets, columns C, R, and S respectively. Satisfying assignment set vector S may be derived by complement of constraint set vector R. Clause set C serves also to define the basis for explicit constraints in the element.

The sets of clauses and constraints are easily represented by use of bit-vectors. Since each clausal partition element is defined by a 3-tuple of variables, taking ($\alpha_1$, $\alpha_2$, $\alpha_3$) to be the abstract representation for such 3-tuple, the possible clauses in C and patterns for constrains on 3-tuples are shown in the left panel of Fig. 1; together with a bit-vector representation of each in the columns headed C and R respectively.

The right panel of Fig. 1 provides an additional illustration depicting the set of admissible assignments as a bit-vector in the column headed S, together with the associated patterns for such in the panel to the right of the column. The bit-vector representation, S, for the admissible assignments can be derived by taking the bit-wise complement of the assignment constraint vector R.

*A. Virtual "Cluster" Structure*

Fig. 2 provides a schematic illustration of the relation of one clausal partition element with all other clausal partition elements in the context of 3SAT. It depicts a virtual "cluster" structure implied for each element within the clausal partition in whole in consequence of variables in common between the elements. Taking $d_i$, the center of the cluster depicted, all other elements of the clausal partition can be partitioned into two sets on basis of having variables in common with $d_i$ or not having variables in common with $d_i$. Those that have variables in common with $d_i$ can then be partitioned into one of the six cells immediately adjacent to $d_i$ in the figure, labeled with ($\alpha_1, \alpha_1\alpha_2, \alpha_2, \alpha_2\alpha_3, \alpha_3, \alpha_1\alpha_3$). The constraints and thus the set of admissible assignments of $d_i$ is fully determined by the both its explicit constraints, in consequence of the clauses within the element, and the implicit constraints imposed by all clausal partition elements in the six cells surrounding it.

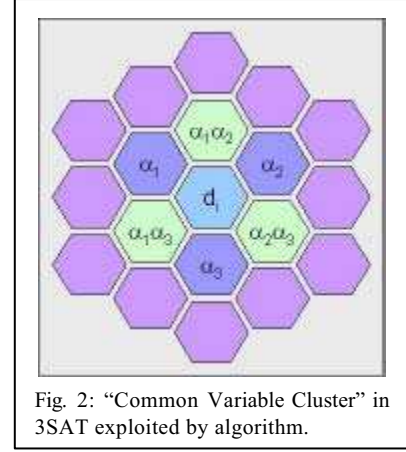

Fig. 2: "Common Variable Cluster" in 3SAT exploited by algorithm.

*B. Properties related to clausal partition elements*

**Proposition 3:** *Necessary & sufficient condition to clausal partition assignment being contained in satisfying assignment of SAT.*

Let $E = (U, C)$ be an instance of a conjunctive expression on some set of disjunctive clauses C in some set of variables U. Let $D = \{d_1, d_2, …, d_n\}$ be the clausal partition of C. If $s \in satset(E)$ and $q \in satset(d_i)$, $d_i \in D$ then $q \subseteq s$ if and only if for all $d_k \in D$ where $(varset(d_k) \cap varset(d_i)) \neq \emptyset$ there exists $p \in satset(varset(d_k) \cap varset(d_i))$ such that $p \subseteq q$ and $p \subseteq r$ where $r \in satset(d_k)$.

**Proof:** (i) [If] – It follows from the definition of satisfiable that if $q \subseteq s$ where s is a satisfying expression, then for all $d_i, d_k \in D$ such that $(varset(d_k) \cap varset(d_i)) \neq \emptyset$ the assignment to variables in $(varset(d_k) \cap varset(d_i))$ is in the set of mutually admissible assignments of $(varset(d_k) \cap varset(d_i))$, and thus in some admissible assignment of both $d_i$ and $d_k$.

(ii) [Only If] – It follows from the definition of satisfiable that $s \in satset(E)$ only if for all $u \in U$ in E, s is an assignment such that E = True. It therefore follows that $q \subseteq s$ only if q is an assignment such that over variables to which q makes assignment, assignment by q to those variables is such that E = True. If thus follows that for all variables to which q makes assignment in an assignment satisfying of E there exists $p \subseteq q$ such that $p \in satset(varset(d_k) \cap varset(d_i))$ $\forall d_k \in D$ where $(varset(d_k) \cap varset(d_i)) \neq \emptyset$.

It follows that the Proposition holds.

**Proposition 4:** *Necessity of mutually admissible assignment to variables in common*

Let $E = (U, C)$ be an instance of a conjunctive expression on some set of disjunctive clauses C in some set of variables U. Let $D = \{d_1, d_2, …, d_n\}$ be the clausal partition of C. Let $d_i$ and $d_j$ be any elements of D where $(varset(d_k) \cap varset(d_i)) \neq \emptyset$. If there exists an assignment satisfying of E then that assignment must contain a mutually admissible assignment to variable(s) in common to $d_i$ and $d_j$, for all $d_i, d_j \in D$.

**Proof:** It follows by Proposition 3. By alternative argument: Assume an assignment satisfying of E exists that does not contain a mutually admissible assignment to variables in common for some $d_i$ and $d_j$. Let $\{a_1, a_2,\ldots, a_n\} = \text{varcom}(d_i, d_j)$. Then for at least one $a_i$ where admissible assignments in $d_i$ require $a_i = x$ and admissible assignments in $d_j$ require $a_i = \neg x$, the assignment to E must take on one of these mutually exclusive assignments. If the assignment assumes $a_i = x$ then at least one clause in $d_j$ cannot be satisfied. If the assignment to E assumes $a_i = \neg x$ then at least one clause in $d_j$ cannot be satisfied. Hence, any conjunction of clauses containing all clauses of C and therefore $d_i$ and $d_j$ cannot be satisfied by such an assignment, a contradiction. Thus, if there exists an assignment satisfying of a conjunction of all clauses of C, then that assignment must contain a mutually admissible assignment to variable(s) in common to $d_i$ and $d_j$, for all $d_i \in D$.

## VI. IMPLICATION OPERATOR

The defined *implication operator* is the principal functional operator of the algorithm. The operator is applied only on elements of the clausal partition in the context of the presented algorithm.

The element from which constraints are being derived is termed the *imposing* set or element. The element receiving the constraints by implication operation is termed the set or element *under imposition*.

Imposition of constraints is always made in terms of the number of variables in the intersection of the variables sets of two elements of the clausal partition on the basis of their respective clause sets. The constraints to be imposed are determined by the assignments that are inadmissible by the imposing element for the variables in common, as a set, in the intersection of the variable sets of the respective imposing element and element under imposition. The imposed constraints are a set of assignments in context of the element under imposition containing assignment to variables in common of values inadmissible by the imposing element.

Constraints that are a consequence of the clauses within an element of the clausal partition are termed *explicit* constraints. It is presumed by the implication operation that all such constraints are imposed at the time of its application. In implementation of the operator it is generally advised that such be explicitly enforced as the initial step of the operator. Constraints imposed on an element are termed *implicit* constraints. No distinction between explicit and implicit constraints is made in the algorithm, however.

### A. Function

The definition of the operator has an implicit assumption that elements of the clausal partition to which it is applied have explicitly imposed constraints on all assignments not satisfying of the clauses of the element. The prior imposition of explicit constraints is necessary to determination of admissible assignments within the element. In implementation of the implication operator this can be enforced using a few logical operations on the bit-vectors for the clause set and constraint set to assure all explicit constraints are imposed during the determination of admissible assignments of an element in the clausal partition.

The implication operator first determines the set of variables in common. It then determines what assignments to the set of variables in common the imposing element of the clausal partition will not admit as admissible assignments, based on the constraints existing in the imposing element. The implication operator then determines which constraints in the element of the clausal partition under imposition contain those inadmissible assignments, and inserts constraints on the determine patterns into the element under imposition. The implication operator thus imposes exclusory constraints from one element of the clausal partition into another, and effectively propagates as consequence the constraints across all the elements of the clausal partition. The consequent effect is that the implication operator constructs the antecedent-consequence chains of implied logical constraints between clausal sets defined by the instance of SAT.

### B. Operator Algorithm

1. Parameters to operator are clausal partition elements A, B; where A is the imposing element and B the element under imposition.
2. Assert all constraints for inadmissible assignments in A.
3. Assert all constraints for inadmissible assignments in B.
4. Determine intersection, {*varset*(A) ∩ *varset*(B)}, of variables defining A and B, respectively.
5. Determine in A the set of inadmissible assignments to {*varset*(A) ∩ *varset*(B)} from step 4.
6. Insert into constraint set of B all assignments containing inadmissible assignments determined for {*varset*(A) ∩ *varset*(B)} in step 5.

### C. Properties

Successive imposition of constraints on a clausal partition element effects a progression from the set {F, T} to the set ∅, of the admissible values assignable to each variable of the element. This progression for individual variables results in a progression in 3-tuples defining of a clausal partition element from ({F, T}, {F, T}, {F, T}) to (∅, ∅, ∅). This progression for 3-tuples occurs in at most eight steps. The progression for a 3-tuple occurs in a path independent manner. A state of (∅, ∅, ∅) is attained only where all possible constraints are imposed on a clausal partition element.

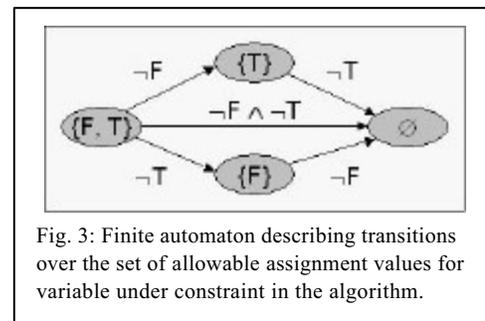

Fig. 3: Finite automaton describing transitions over the set of allowable assignment values for variable under constraint in the algorithm.

The finite automaton describing the progression in reduction of the set of admissible value assignments for a variable, based on exclusion constraints is given in Fig. 3. Progressive reduction of admissible assignments for the 3-tuple of variables defining of clausal partition elements is definable as a 3-tuple of finite automata presented in Fig. 3, where each is an independent finite automata. The generation of the finite automaton for clausal partition elements is given in appendix, with the full finite automata given in tables attached to this paper.

The algorithm's success is predicated on the behavior defined in the finite automaton for admissible values assignments of individual variables. The behavior provides the necessary basis for the reductive property of the implication operator on the admissible assignments in each element of the clausal partition. (Non-reductive on constraints.) The behavior assures that where a value is excluded from admissible value assignments for a variable it cannot reappear; thus a reduction of the set of constraints in an element of the clausal partition subject to subsequent implication operations is precluded. This assures that the admissible assignment set in clausal partition elements is reductive under application of the implication operator. Since by definition the implication operator not remove any constraint once established, that the admissible assignment set of clausal partition elements adhere with respect to individual variables to the transition scheme defined by the finite automaton given in Fig. 3 is also assured as an inherent consequence.

**Proposition 5:** *Implication Operator $\nabla$ is reductive.*

**Proof:**  Given any instance of A$\nabla$ B, by definition the implication operator imposes constraint from A into B such that the imposed constraints in B eliminates from the satset(B) assignments inconsistent with the constraints. The implication operator therefore causes a reduction of zero or more admissible assignments in satset(B). Thus, by definition of reductive property the implication operator is reductive on the set of admissible assignments.

**Proposition 6:** *Implication Operator $\nabla$ is not strictly reductive.*

**Proof:**  By Proposition 5 the implication operator is reductive. Given any case where the common variables under the implication operator are consistent between sets on which the operators is applied prior to such application, it follows that zero elements from the admissible assignments of the set on to which constraint is imposed will be excluded. Thus, by definition the implication operator is not strictly reductive on the set of admissible assignments.

**Proposition 7:** *Implication Operator $\nabla$ is not non-reductive.*

**Proof:**  Given any instance of A$\nabla$B, it follows by definition of the implication operator that such cannot remove an existing constraint; thus is not non-reductive on the set of admissible assignments in B.

## VII. INSTANCE GRAPH

The instance graph forms the primary structure over which execution of the algorithm is performed. The instance graph encodes the representation of the common variable relationship between elements of the clausal partition. It thus provides an object representation of an instance of 3SAT being solved by the algorithm. The vertices of the graph are associated with the elements of the clausal partition. The edges of the graph represent the existence of a common variable relationship between elements of the clausal partition. The algorithm iteratively applies the implication operator to the pairs of elements in the clausal partition defined by the edges of the instance graph.

**Proposition 8:** *3SAT Cardinality V in Instance Graph is $O(|U|^3)$*

Let E = (U, C) be an instance of 3SAT. Let G = (V, $\beta$) be an instance graph of E. The maximum cardinality of V is $O(n^3)$, where n = |U|.

**Proof:**  Given any set of 3 objects drawn from n = |U| items there are at most $n^3$ such sets. It follows by definition of the clausal partition, D, of C in E that the cardinality of D is at most $n^3$. Given that V is defined by a one to one and onto relation with D, the cardinality of V is therefore at most $n^3 \approx O(n^3)$.

**Proposition 9:** *3SAT Degree of all v∈V of Instance Graph is $O(|U|^2)$*

Let E = (U, C) be an instance of *3SAT*. Let D be the clausal partition of E. Let G = (V, β) an instance graph of E. $\forall$ $v_i \in$ V, degree of $v_i$ is at most $O(n^2)$, where n = |U|, the number of variables in E.

**Proof:** By definition of G an edge exists in β only where two vertices, $v_i$ and $v_j$, to which the edge is incident, are associated with elements, $d_i$, $d_j \in$ D, such that 0 < |varset($d_i$) ∩ varset($d_j$)| < 3. Let $a_1$, $a_2$, $a_3$ be the variables of $d_i \rightarrow v_i$, then edges in G incident to $v_i$ define a common variable relation between $v_i$ and other vertices in G on basis of the power set over the variables of $d_i$ (i.e.: {{$a_1$}, {$a_2$}, {$a_3$}, {$a_1$, $a_2$}, {$a_1$, $a_3$}, {$a_2$, $a_3$}}). Therefore, for {{$a_1$}, {$a_2$}, {$a_3$}} there can exist at most $3(n)(n) = 3n^2$ possible combinations of variables not in the set of clauses defining $d_i$. Similarly, for {{$a_1$, $a_2$}, {$a_1$, $a_3$}, {$a_2$, $a_3$}} there can exist at most 3(n) possible combinations of variables not in the set of clauses defining $d_i$. Consequently, $v_i$ can have at most $(3n^2 + 3n) \approx O(n^2)$ incident edges in G, where n = |U|, the number of variables in E.

**Proposition 10:** *3SAT Execution of Implication Operation over all v ∈ V in Instance Graph is $O(n^5)$*

Let E = (U, C) be an instance of 3SAT. Let D be the clausal partition of E. Let G = (V, β) be the instance graph of E. At most $O(n^5)$ time, where n = |U|, the number of variables in E, is required to perform once all instances of the implications $v_i \nabla v_j$ defined in G by β.

**Proof:** The implication operation $\nabla$ requires constant order time in the cardinality of U to be performed. Since there are n = |U| variables in E there are at most $n^3$ distinct elements of D and therefore $|V| \leq n^3$. By Proposition 9 for all v ∈ V degree(v) ≤ $O(n^2)$. It follows then that for any $v_i \in$ V the resolution of $v_i \nabla v_j$ for all $v_j \in$ V such that $(v_i, v_j) \in$ β requires then at most $O(n^2)$ distinct executions of the implication operation $\nabla$, where n = |U|. By Proposition 8 $|V| \leq O(n^3)$. Thus, $(O(n^2)O(n^3)) \rightarrow O(n^5)$ distinct executions of the implication operation are necessary to perform once all instances of implication defined in G by β, where n = |U|, the number of variables in E.

## VIII.  ALGORITHM

The algorithm is primarily an iterative application of the implication operator on the clausal partition of an instance of 3SAT. The algorithm applies the implication operator in iterative manner to propagate the consequence of all existent constraints in the system until a steady state is attained, or, alternatively, an iterative limit is exceeded.

*A. Description*

The algorithm uses the clausal partition and instance graph to define a network representation of the instance of SAT that the algorithm then uses to apply the implication operator in succession to pairs of elements in the clausal partition until a steady state is reached. The successive application of the implication operator effectively determines the necessary agreement of clausal partitions respective admissible value assignments to variables such hold in common in the context of the SAT instance in whole.

The effect of the implication operation over the clausal partition in the instance graph is the encoding of the set of all inadmissible assignments as constraints in the context of SAT instance – effectively an encoding of the set of non-satisfying assignments. The algorithm thus derives by compliment an answer to the question of existence of a non-empty set of assignments satisfying of the SAT instance.

The algorithm applies the implication operator outward from the center cell of the cluster structure depicted in Fig. 2, for each element of the clausal partition. The net effect of such is that constraints of elements in each of the six adjacent cells flow into $d_i$, in consequence of each of the elements in turn being in either the center cell or one of the immediately adjacent cells of such a cluster with every element of the clausal partition with which it has variables in common.

The fact that all clausal partition elements in the six cells immediately adjacent to it are under imposition of those existing in cells beyond them leads to $d_i$ by consequence of the imposed constraints on elements occupying the adjacent cells, also being subject to the consequence of constraints of distant elements of the clausal partition of E.

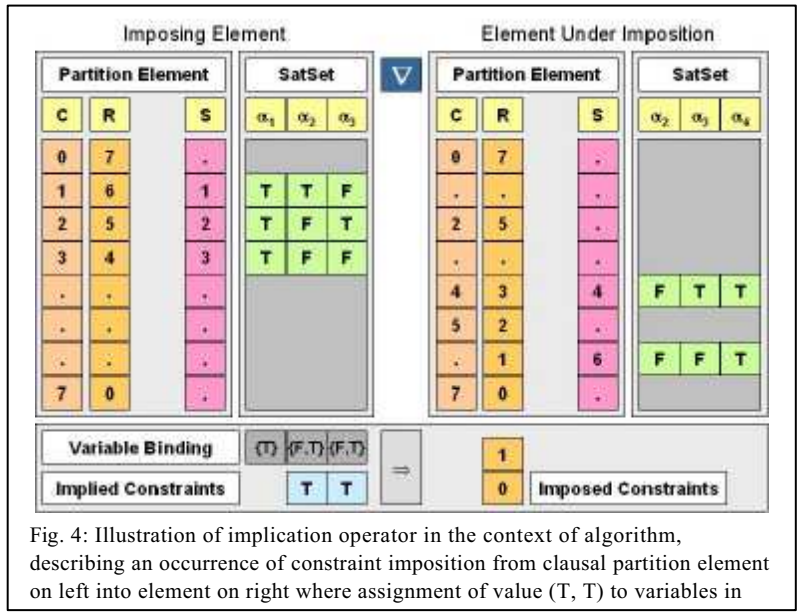

Fig. 4: Illustration of implication operator in the context of algorithm, describing an occurrence of constraint imposition from clausal partition element on left into element on right where assignment of value (T, T) to variables in

Fig. 4 is illustrative of the action of the implication operator, in the context of the algorithm. The implication is being illustrated with the left instance imposing constraint onto the right instance. The left case, having all four instances of clauses having $\alpha_1$ not negated, cannot allow $\alpha_1 = F$. It therefore must bind $\alpha_1 = T$ and evaluate for constraints on the variables in common with the instance on the right of the illustration (i.e.: $\alpha_2$ and $\alpha_3$). Had the instance on the left not included clauses for $\alpha_1 = F$ it would not care what values were ever assigned to $\alpha_2$ and $\alpha_3$; thus would not impose any constraint upon these variables. However, with clause pattern number 7 present in the set, it cannot admit a satisfying assignment with any clause set with which it shares these variables unless the variables do not take on the values {T, T}, respectively, as binding forces $\alpha_1 = T$. The consequence to the instance on the right, under imposition, is to add to its constraint set constraints {0, 1}. Since both constraints exist none would be added to the constraint set of the right hand instance.

*B. Algorithm*

The algorithm is of two parts. The parts are identified as "Part A" and "Part B". *Part A* determines the set of all satisfying assignments for an instance of SAT, and thus provides an answer to the question of existence of a satisfying assignment. *Part B*, making use of the results of *Part A*, is then able to determine a specific instance of a satisfying assignment for an instance of SAT.

**Part A:** (Answering: "Does E have a non-empty set of satisfying assignments?")
1. Let $E = (U, C)$ be an instance of SAT in 3SAT form.
2. Partition C into a clausal partition $D = \{d_1, d_2, \ldots d_m\}$.
3. Define for E an instance graph $G = (V, \beta)$.
4. For all $v_i \in V$ determine initial constraints of associated $d_i \in D$.
5. For all $v_i \in V$, for each $(v_i, v_j) \in \beta$, apply implication operator, $v_i \nabla v_j$. The consequence is a reduction of admissible assignments in $d_j$ associated with the adjacent vertex $v_j$ by any having values to varset($d_i, d_j$) inconsistent with new constraints imposed.
6. Repeat step 5 until a steady state is attained.
7. If a steady state is reached such that for all $v_i \in V$ the associated set of clauses, $d_i$, has a non-empty set of satisfying assignments, then E has at least one solution for which it evaluates to True, otherwise E has no solutions for which it evaluates to True.

**Part B**: (Answering: "An assignment for which E evaluates to True?")

8. Where E has at least one solution by Step 7 above, For $k = 1\ldots|V|$:
   a. Select one admissible assignment in $d_i \in D$ and impose constraint against all others.
   b. Perform step 5, starting with $v_i = v_k$, as selected here.
   c. Repeat step 8b until a steady state is attained.
9. G contains in V the assignment of variables for one instance of a solution of E.

*Part A* attains a steady state condition that encodes in the structure of the network the set of assignments not satisfying of the SAT instance, and by complement of that set the set of all assignments satisfying of the instance. If no clausal partition element has at steady state all possible constraints imposed, implying an empty set of admissible assignments, then the SAT instance has at least one satisfying assignment. If there exists an assignment satisfying of the SAT instance by the results of *Part A*, then *Part B* when applied to those prior results, can extract from such an instance of an assignment satisfying to the SAT instance.

C. *Completeness and Correctness of Algorithm Part A*

**Theorem 1:** *3SAT Steady state of algorithm does not contain as admissible assignment in any clausal partition element any that are not also admissible in at least one assignment satisfying of SAT instance.*

Let $E = (U, C)$ be an instance of 3SAT. Let D be a clausal partition on C. Let $G = (V, \beta)$ be an instance graph defined on D for E. *Part A* of the algorithm, at steady state, does not contain in any clausal partition element admissible assignments not also admissible in at least one assignment satisfying of the instance of 3SAT.

   **Proof:**   Assume there exists a steady state $s_l$ in which there exists a clausal partition element that contains as admissible an assignment that is not admissible in at least one assignment satisfying of E. This implies that there exists either: (1) a $d_i \in D$ that at steady state admits as an admissible assignment that does not satisfy all clauses in $d_i$, or; (2) a $d_i \in D$ that at steady state admits as an admissible assignment one that does not also admit a mutually admissible assignment to variables in common with at least one $d_k \in D$, as required by Proposition 3 and Proposition 4. Case (1) contradicts the definition of satisfiable and the initial (explicit) constraint imposition in each element of the clausal partition of only those assignments non-satisfying of the clauses in the element independent of context. Case (2) contradicts the definition of both steady state and the implication operator. Such infers a failure to impose constraint where assignment to variables in common was not admitted by all elements of the clausal partition containing the variables. Thus, it follows that the algorithm at steady state does not contain as admissible assignments for any element in the clausal partition assignments not also admissible in at least one assignment satisfying of E, the instance of SAT.

**Theorem 2:** *3SAT Steady state of algorithm contains as admissible assignment in each clausal partition elements all such that are admissible in at least one assignment satisfying of SAT instance.*

Let E = (U, C) be an instance of 3SAT. Let D be a clausal partition on C. Let G = (V, β) be an instance graph defined on D for E. *Part A* of the algorithm, at steady state, contains as admissible assignment in each clausal partition element all assignments to the 3-tuple of variables defining of the clausal partition element that are admitted by at least one assignment satisfying of the instance of SAT, and therefore contains all assignments satisfying of the instance.

**Proof:** Assume there exists an assignment satisfying of E that at steady state $s_2$ is not contain in the result of algorithm *Part A*. This implies that for at least one $d_i \in D$ there exists an admissible assignment to the variables of $d_i$ that is not an admitted by the results as an admissible assignment by the set of constraints in $d_i$. Existence of the assignment to variables of $d_i$ in an assignment satisfying of E implies that such assignment is (1) satisfying of all clauses in $d_i$, and; (2) admitting of a mutually admissible assignment for all $d_k \in D$ with which $d_i$ has variables in common, as required by Proposition 3 and Proposition 4. Case (1) contradicts the definition of satisfiable and the initial (explicit) constraint imposition in each element of the clausal partition of only those assignments non-satisfying of the clauses in the element independent of context of chain. Case (2) contradicts the definition of the implication operator by inference of an imposition of constraint where assignment to variables in common was admitted by all elements of the clausal partition containing the variables. Thus, it follows that the algorithm at steady state contains all admissible assignments for all clauses in the context of E, and therefore contains all assignments satisfying of E.

**Theorem 3:** *3SAT Algorithm steady state is unique*

Let E = (U, C) be an instance of 3SAT. Let D be a clausal partition on C. Let G = (V, β) be an instance graph of E. The steady state attained by the algorithm of *Part A* is unique.

**Proof:** Assume two different steady states, $s_1$ and $s_2$, were to be attainable. This implies that (1) $s_1$

contains at least one assignment not contained in $s_2$, or; (2) $s_2$ contains at least one not contained in $s_1$, or; (3) both condition (1) and (2) hold. Either case implies that either (a) at least one of the steady states, $s_1$ or $s_2$, contains at least one assignment that is an inadmissible assignment, or; (b) at least one of the steady states, $s_1$ or $s_2$, does not contain at least one admissible assignment. Case (a) contradicts Theorem 1, while case (b) contradicts Theorem 2. Thus, by contradiction, it follows that steady state resolution of the algorithm is unique.

### D. Completeness and Correctness of Algorithm Part B

**Theorem 4**: *3SAT Algorithm Part B determines an instance of solution of SAT where Part A determines existence of a non-empty set of satisfying assignments.*

**Proof:** Given *Part A* determines the existence of a non-empty set of assignments satisfying of an instance of SAT, each clausal partition element possesses a set of constraints such that only those assignments that may exist within at least one assignment satisfying of the SAT instance are admissible, by Theorem 1, Theorem 2, and consequence of Theorem 3. Assume that an assignment satisfying of the instance of SAT is not obtained in step 9, it follows in consequence of the correctness of *Part A* and by proof the prior theorems that the assumption results in a contradiction; therefore *Part B* by successive constraint on the resultant set of admissible values reduces the set of all assignments that encoded by result of *Part A* to one assignment satisfying of an instance of SAT.

## IX. COMPLEXITY

### A. Steady State Time Complexity of Part A

**Definition:** 1-*Operator Discrete Temporal System*

$S = (D, \beta, M, R, \delta, n, t)$. $D$ is an n-tuple, the elements of which take their values in the set of positive integers. $\beta$ is a set of unordered pairs, where each pair is defining of a relation in S between two elements of D. M is a set of ordered 3-tuples $m = (d_i, \alpha, \omega)$, defining limits on the range of values for each $d_i \in D$ as such that $\alpha \leq d_i \leq \omega$. R is a set of transforms invoked by temporal operator. $\delta$ is the system temporal operator. *n* is a defined constant of the system and the cardinality of D. *t* is the system temporal index. S is thus defined to be a 1-Operator Discrete Temporal System.

**Definition:** *Instance of $S = (D, \beta, M, R, \delta, n, t)$ to be used by Proposition and Theorem of paper.*

Let S defined a 1-Operator Discrete Temporal System, where: $D = (d_1, d_2, …, d_n)$; $\beta = \{(d_i, d_j) \mid$ where a relation exists between $d_i, d_j \in D\}$; $M = \{m = (d_i, \alpha, \omega) \mid d_i \in D, \alpha, \omega \in Z\}$; $R = \{r \mid r \to d_i = d_i + q, q < 0$ where $t_{n-1}(d_i) > t_n(d_i)$, otherwise $q = 0 \}$; $\delta = \{\forall d_i$ where $\exists (d_i, d_j) = e \in \beta \Rightarrow r(d_i)\}$; t has initial value $t = 0$.

**Proposition 11:** *S is a reductive system.*

**Proof:** By definition of R there exists no $r \in R$ such that for any instance of $\delta$ the consequence of $r(d_i) \Rightarrow d_i = d_i + q$ where $q > 0$, it follows in consequence that $\forall d_i \in D$ $d_i$ cannot increase in value. Thus, S is reductive and describes a monotonically decreasing sequence in the value of each $d_i \in D$.

**Proposition 12:** *S attains a steady state*.

**Proof:** By Proposition 11 S defines a reductive temporal system wherein δ implies a monotonically decreasing sequence in value of the elements of D, then either: (i) by consequence of bounds on the values for all $d_i \in D$ in the context of S defined by M all $d_i$ attain their lower bounds; or, (ii) for some value of $t = f$ δ applies R such that $\forall d_i \in D\ r(d_i) \Rightarrow d_i = d_i$. It follows in consequence of (ii) that for all values of $t > f$ δ applies R such that $\forall d_i \in D\ r(d_i) \Rightarrow d_i = d_i$. Thus, it follows that S attains a steady state.

**Proposition 13:** *S attains a steady state in time polynomial in initial conditions*.

**Proof:** It follows from Proposition 12 there must exist at least one $d_i \in D$ such that $d_i$ is reduced in each time step if the system is not to enter a steady state. Therefore, by consequence of definition of R in context of S it follows that at least one $d_i \in D$ must be reduced in each time step by 1. It follows then that such reductions can occur for at most $t = \left( \sum_{i=0}^{n} | m(d_i, \omega) - m(d_i, \alpha) | \right)$ steps in S. Thus, S attains a steady state in polynomial time in the initial conditions.

**Theorem 5:** *3SAT Algorithm Part A is polynomial time reducible to S*.

**Proof:** Let $E = (U, C)$ be an instance of 3SAT. Let $G = (V, \beta)$ be an instance graph of E. Let D in the context of S be defined as an n-tuple for which the values are the cardinality of the set of admissible assignment to corresponding $d_i$ in the clausal partition of C. Let β in the context of S be defined by β in the context of G. Let $M = \{m = (d_i, \alpha, \omega) \mid \forall\ d_i \in D\ \alpha = 0,\ \omega = 8)\}$. Let $R \approx \nabla$. Let $n = |V|$. Let $t = 0$. Since each assignment is polynomial time achievable it follows the algorithm of *Part A* is polynomial time reducible to S.

**Theorem 6:** *3SAT Steady state time complexity in S is polynomial in the cardinality of U*.

Let $E = (U, C)$ be an instance of 3SAT. Let S' be defined as given in Theorem 5 for an instance of SAT. S' attains steady state in time polynomial in the cardinality of U, the variables of the SAT instance.

**Proof:** The implication operator being reductive on the cardinality of the set of admissible assignment by Proposition 5, it admits equivalent requirement of δ in S' that $R \Rightarrow \{r(d_i)_{n-1} \geq r(d_i)_n\}$; therefore S' is an equivalently reductive system. It follows from definition of the clausal partition on C and of 3SAT that $|D| \leq O(|U|^3)$. For all $d_i \in D$ the cardinality of the set of admissible assignments for $d_i$ is bounded such that $0 \leq |satset(d_i)| \leq 8$, by the definition of D and of 3SAT. It follows then by Proposition 13:

$$\left( \sum_{i=0}^{n} | m(d_i, \omega) - m(d_i, \alpha) | \right) \leq (8 * |U|^3) \approx O(|U|^3).$$

Proposition 9 establishes that for each $v_i \in V$ the degree of $v_i$ is less than or equal to $O(|U|^2)$; by definition of $v_i \in V$ on basis of the clausal partition of C, it follows that δ must evaluate $d_i \nabla d_k$ at most $O(|U|^2)$ times to determine and communicate the implication of constraints from each $d_i$ to all $d_k$ with which there may exist an edge in β context of G, hence an ordered pair in β in context of S, for each $d_i \in D$. For each time step, thus each iteration of step 5, δ must therefore evaluate $\nabla$ a total of $O(|D| * |U^2|)$ times, which is equivalent to $(O(|U|^3) * O(|U|^2)) \approx O(|U|^5)$. Since the time order of S is $O(|U|^3)$ steps wherein $\delta$ is evaluated once in each step, it follows that time order of S' in terms of U is such that $t \leq (O(|U|^3) * O(|U|^5)) \approx O(|U|^8)$ at steady state. Since steady state can be determined by flagging the evaluation of $r(d_i)$, such adds at most a constant to the order of t at steady state. Thus, S attains a steady state in time polynomial in the initial conditions in terms of U, the cardinality of set of variables in an instance of 3SAT and, therefore has polynomial time complexity in the cardinality of U.

## B. Time Complexity Part B

**Theorem 7:** *3SAT Solution determinable in polynomial time*

Let E = (U, C) be an instance of *3SAT*. The existence of a solution of E can be determined in polynomial time in the number of variables in E, and; where a solution exist a solution can be determined in polynomial time in the number of variables in E.

**Proof:** By Theorem 5 the question of existence of a solution to E is attainable in polynomial time in the number of variables in E by *Part A* of the algorithm. By Theorem 4, algorithm *Part B* derives an instance of an assignment satisfying of E. Without loss of generality or effectiveness, assume G is a connected graph, as each disconnected subgraph in G may otherwise be considered independently without detriment to proof. The repeated process of selection and reduction in *Part B* requires one execution of *Part A* for each vertex in G. *Part B* then has time complexity in worst case of $(O(n^3)(On^8)) \to O(n^{11})$. Thus, determination of an assignment satisfying of 3SAT is polynomial in |U|, the number of variables in E.

## C. Space Complexity

**Proposition 14:** *Necessary & sufficient conditions to determination of admissible assignments of clausal partition elements.*

Let E = (U, C) be an instance of 3SAT. Let D be a clausal partition on C. For all $d_i \in D$, $satset(d_k)$ for all $d_k \in D$, such that for $(varset(d_i) \cap varset(d_k)) \neq \emptyset$, is necessary and sufficient to determination of $satset(d_i)$.

**Proof:** It follows from Proposition 3 that for any $q \in satset(d_i)$, $satset(d_k)$ for all $d_k \in D$, such that for $(varset(d_i) \cap varset(d_k)) \neq \emptyset$, is necessary and sufficient to determination of $q \in satset(d_i)$. It follows by consequence that $satset(d_k)$ for all $d_k \in D$, such that for $(varset(d_i) \cap varset(d_k)) \neq \emptyset$, is necessary and sufficient to determination of all $q \in satset(d_i)$; and thus to determination of $satset(d_i)$, the admissible assignments of clausal partition elements.

**Theorem 8:** *Representation of constraint set of $d_i$ requires at most 8 bits for each $d_i \in D$.*

**Proof:** Proposition 14 establishes that the consequence of constraint upon $(varset(d_i) \cap varset(d_k))$ for all $d_k \in D$ is necessary and sufficient to determination of $satset(d_i)$. Proposition 3 establishes that if there exists $q \in asgspc(d_i)$ such that for some $d_k \in D$ where $(varset(d_k) \cap varset(d_i)) \neq \emptyset$ there does not exist at least one $p \in satset(varset(d_k) \cap varset(d_i))$ such that $p \subseteq q$ and $p \subseteq r$ where $r \in satset(d_k)$, then for all $q \in asgspc(d_i)$ such that $p \subseteq q$ it follows that $q \notin satset(d_i)$. It is therefore sufficient that a set of exclusory constraints upon the set of all assignments of $d_i$ necessary to indicate $satset(d_i)$ need only represent the set of excluded assignments in asgspc($d_i$). Since the asgspc($d_i$) ≤ 8 for all $d_i \in D$, it follows that at most 8 bits is required for representation of constraint set of $d_i$.

**Theorem 9:** *3SAT Space complexity is polynomial in the size of U (SAT variable set).*

**Proof:**　　Let E = (U, C) be an instance of 3SAT. Let D be a clausal partition on C. Let G = (V, β) be an instance graph of E. It is well known that a graph, such as G, has space complexity polynomial in $|V| \approx O(|\log_2 V|^2)$. Since $|V| \approx O(|U|^3)$ by Proposition 8, the space complexity of G is $O(\log_2 |U|^6)$. The data required by system is, for all $d_i \in D$, the clausal set and constraint set of $d_i$. By definition of instance graph $|D| = |V| \approx O(|U|^3)$. By Theorem 8 at most 8 bits are required per instance of $d_i \in D$ for the representation of the constraints upon $d_i$. From the definition of E it follows that for each $d_i \in D$ there exists at most 8 possible assignments for each $d_i$, thus such can be represented with at most 8 bits per $d_i \in D$. Where U is explicitly represented such requires $O(\log_2 |U|)$ bits. It follows that the algorithm has space complexity give by:

$$(O(\log_2 |U|^6) + O(|U|^3) + O(\log_2 |U|)) \to O(|U|^3) \approx O(n^3),$$

where $|U| = n$. The space complexity of the algorithm is therefore polynomial in the size of the variable set in 3SAT.

X.　Polynomial Time Solution of SAT

**Theorem 10:** *3SAT Existence of a solution to SAT is determinable in time polynomial in the number of variables in an instance of SAT, and is determined by algorithm Part A.*

**Proof:**　　Let E = (U, C) be an instance of 3SAT. Let D be the clausal partition of C. Let G be the instance graph of E defined on D. Theorem 6 establishes that the algorithm of *Part A* must determine a steady state within a polynomial time bound in the cardinality of |U|, the number of variables in the SAT instance. Theorem 1 establishes that at steady state the clausal partition elements contain as admissible only those assignments to the 3-tuple of variables defining of each element that are also admissible in at least one assignment satisfying of the instance of 3SAT. Theorem 2 establishes that at steady state the clausal partition elements contain as admissible all assignments to the 3-tuple of variables defining of each element that are admissible in at least one assignment satisfying of the instance of 3SAT. Theorem 3 establishes that the steady state attained by the algorithm is unique. It follows then that existence of an assignment satisfying of an instance of SAT is determinable in time polynomial in the number of variables in an instance of SAT, and that algorithm *Part A* determines existence of an assignment satisfying of the instance of SAT.

**Theorem 11:** *3SAT An instance of an assignment satisfying of SAT is determinable in time polynomial in the number of variables in an instance of SAT, and is determined by algorithm Part B.*

**Proof:**　　Let E = (U, C) be an instance of 3SAT. Let D be the clausal partition of C. Let G be the instance graph of E defined on D. Theorem 12 establishes that the algorithm of *Part A* determines the existence of an assignment satisfying of an instance of SAT. Theorem 4 establishes that where *Part A* determines existence of an instance of an assignment satisfying of an instance of SAT, then algorithm *Part B* determines an instance of such an assignment. Theorem 7 establishes that *Part B* determines an instance of such an assignment in time polynomial in the cardinality of U, the number of variables in the instance of SAT. It follows then that an instance of an assignment satisfying of an instance of SAT is determinable in time polynomial in the number of variables in an instance of SAT, and that algorithm *Part B* determines an assignment satisfying of an instance of SAT.

## XI. SUMMARY

### A. Of Algorithm

The algorithm presented derives a partition of the set of all possible assignments for any given instance of SAT into the set of assignments not satisfying of the instance and the set assignments satisfying of the instance. The algorithm performs this derivation of the partition through an encoding of the set of inadmissible assignments to a partition, on basis of 3-tuples of variables in the instance, of the sets of clauses in the instance of SAT, and thus by compliment the set of admissible assignment. The encoding is itself accomplished by means of evaluation of a defined operator propagating implied constraints on variables in common between elements of a clausal partition defined on the set of all clauses in the given instance of SAT, so as to exclude by constraint from the set of all admissible assignments those assignments in each clausal partition element that either (1) fail to satisfy the clauses of the set or (2) fail to admit at least one mutually admissible assignment to variables in common with any other clausal partition elements. The algorithm thus makes effective use of the set of non-satisfying assignments of the instance of SAT to reduce the set of all assignments to a set containing only assignments satisfying of the instance. The algorithm then reduces the set of all assignment satisfying of the instance to an instance of an assignment satisfying of the instance of SAT.

### B. Of Algorithm Vulnerabilities

The algorithm can be shown to fail only where it can be shown:

1. To fail to be reductive on the set of admissible assignments within each element of the clausal partition, or;
2. To fail to exclude as admissible from at least one element of the clausal partition some assignment that cannot exist in any assignment satisfying of some instance of SAT, or;
3. To fail to not exclude as admissible from at least one element of the clausal partition some assignment that may exist in at least one assignment satisfying of some instance of SAT.

Case 1 requires that an admissible assignment once exclude by readmitted, contradicting there having existed cause to have excluded the assignment, in contradiction of the definition of the implication operator.

Case 2 requires that an assignment to some element in the clausal partition be admissible that is either not satisfying of the clauses in the set contained in the element, or does not admit a mutually satisfying assignment to variables in common with at least one other element of the clausal partition, contradicting the definition of the implication operator and the system being in steady state.

Case 3 requires that an assignment to some element in the clausal partition be exclude that is both satisfying of the clauses in the set contained in the element, and admits a mutually satisfying assignment to variables in common with all other elements of the clausal partition, contradicting the definition of the implication operator.

## C. Of Future Work

The worst case time complexity for the algorithm present is shown to be $O(n^8)$. It is conjectured that average run time will likely be a far more reasonable $O(n^2)$ to $O(n^3)$. An average case time complexity for the algorithm is subject of current research we hope to address in future paper.

<div align="center">APPENDIX</div>

## A. Clausal Partition Element Constraint State Machine

The state machine for clausal partition element progression from no constraints to maximal constraints is produced by state transitions on basis of a bit being set, representing insertion of additional constraint into the set. Since there are 8 possible constraints and 256 possible combinations of them, the transition table is 256 x 8 with one column for each bit. The cells of the table then contain the binary value of the current state binary-OR'd with the bit being set.

The transitions of the state machine for each variable, as described in section VI, occur in relation to the absence of all four assignments containing the value. The T/F association with the clausal partition element pattern set for each variable, based on the prototypical model for clausal partition elements depicted in Fig. 1, is as follows, least significant bit to the right:

$$\alpha_1 : (T = 00001111, F = 11110000)$$
$$\alpha_2 : (T = 00110011, F = 11001100)$$
$$\alpha_3 : (T = 01010101, F = 10101010)$$

## B. Clausal Partition Element Admissible Assignments by Common Variables

The generation of the inadmissible assignment set (admissible assignment set) for each of the six common variable patterns $\{\alpha_1, \alpha_2, \alpha_3, \alpha_1\alpha_2, \alpha_2\alpha_3, \alpha_1\alpha_3\}$ is generated based on presence (absence) of all occurrences of the respective patterns for the variable subset in the constraint (satisfying) set. The result is either: $\varnothing$, where no assignment is possible due to one or both variables being fully constrained; or, a subset of $\{\{T\}, \{F\}\}$, in the case of $\alpha_1, \alpha_2, \alpha_3$, or, $\{\{T, T\}, \{T, F\}, \{F, T\}, \{F, F\}\}$ in the case of $\alpha_1\alpha_2, \alpha_2\alpha_3, \alpha_1\alpha_3$.

For example: Where the constraint state is 47, the corresponding binary value is "00101111" (least significant bit to the right). Taking the variables $\alpha_1\alpha_2$, the binary value in combination with the T/F association patterns (see above) indicates that for $\alpha_1$ all patterns constraining T are present, while for $\alpha_2$ at least one pattern is present for both T and F. Consequently, the inadmissible assignments for $\alpha_1\alpha_2$ would be $\{\{T, T\}, \{T, F\}\}$, and the admissible assignments $\{\{F, T\}, \{F, F\}\}$. The process can be readily accomplished using bit level operations.

## CORRECTIONS

The following grammatical corrections were made:

1. Section I, Paragraph 1: Removed quotes around question text in first sentence; completed reference to [2] in second sentence; inserted word "also" into last sentence.
2. Section I, Paragraph 2: Changed "where in" to "wherein" in line three.
3. Section II, Paragraph 3: Added reference to [2] in line two.
4. Section II, Paragraph 4: Inserted "to" before "individual" in last line.
5. Section II, Paragraph 6: Inserted "}" to close clausal partition set in example.
6. Section II, Paragraph 8: Changed "containing" to "contained" in line three.
7. Section II, Paragraph 9: Changed "of" to "on" in line one; changed "exclusive" to "exclusory" in line two; removed "of" before "to a variable" in line four.
8. Section II, Paragraph 13: Inserted "on the" between "8" and "number".
9. Section V, Paragraph 4: Removed comma in line one.
10. Section VI, Paragraph 1: Inserted "of the variable sets" after "intersection".